\documentclass[aps,prl,twocolumn,groupedaddress,showpacs]{revtex4}

\usepackage{amsmath}
\usepackage{graphicx}
\usepackage{bm}
\usepackage{hyperref}

\newcommand{\Hca}{\mathcal{H}}

\newcommand{\rA}{\text{A}}
\newcommand{\rB}{\text{B}}

\newcommand{\tp}{t_{\perp}}

\newcommand{\e}{\epsilon}
\newcommand{\Eg}{E_{\text{g}}}
\newcommand{\kg}{k_{\text{g}}}
\newcommand{\mg}{m_{\text{g}}}
\newcommand{\Eb}{E_{\text{b}}}

\newcommand{\om}{\omega}

\newcommand{\simleq}{\mbox{\raisebox{-1.0ex}{$\stackrel{<}{\sim}$}}}

\newcommand{\vk}{{\bf k}}

\newcommand{\vf}{v_{\rm F}}

\newcommand{\Ima}{{\rm Im}}

\newcommand{\GR}{{\rm G}}
\newcommand{\PD}{\partial}

\begin{document}

\title{Impurities in a Biased Graphene Bilayer}

\author{Johan Nilsson}
\affiliation{Department of Physics, Boston University, 590 
Commonwealth Avenue, Boston, MA 02215}

\author{A.~H. Castro Neto}
\affiliation{Department of Physics, Boston University, 590 
Commonwealth Avenue, Boston, MA 02215}

\date{\today}

\begin{abstract}
We study the problem of impurities and mid-gap states in a biased graphene
bilayer. We show that the properties of the bound states, such 
as localization lengths and binding energies, can be controlled 
externally by an electric field effect. Moreover, the band gap is
renormalized and impurity bands are created  at finite impurity
concentrations. Using the coherent potential approximation we
calculate the electronic density of states and its dependence on the
applied bias voltage. 
\end{abstract}

% insert suggested PACS numbers in braces on next line
%\keywords{}

\maketitle

The discovery of single layer graphene \cite{Netal04_short}, and its further experimental
characterization demonstrating unconventional metallic properties
\cite{Novolelov2005_short,Zhang2005_short}, have attracted a lot of attention 
in the condensed matter community. The graphene bilayer, which is
made out of two stacked graphene planes, is a particularly
interesting form of two-dimensional (2D) carbon because of
its unusual physics that has its origins on the peculiar band structure.
At low energies and long wavelengths it can be described in terms of
massive, chiral, Dirac particles.
Previous studies of the graphene bilayer have focused on
the integer quantum Hall effect \cite{Falko2006a,Novoselov2006_bilayer_short}, 
the effect of electron-electron interactions \cite{Nilsson06a_short}, and 
transport properties 
\cite{Nilsson06b_short,Ando06a,Katsnelson06c,Snyman06,Falko2006b}.
An important property of this system is that when an electric field is
applied between the two graphene layers, an electronic gap opens in
the spectrum \cite{Falko2006a,McCann06,Paco06a}; 
the resulting system is called the 
{\it biased graphene bilayer} (BGB).
In contrast, the single layer graphene does not have this unique
property since to generate a gap in the electronic spectrum the
sub-lattice symmetry of the honeycomb lattice has to broken,
and this is a costly energetic process. 

Ohta {\it et al.} have studied graphene films on SiC substrates using
angle-resolved photoemission spectroscopy (ARPES) \cite{Ohta06_short},
and spectra  
reminiscent of the BGB were uncovered. Nevertheless, the graphene films are 
heavily doped by the SiC and the chemical potential and gap value cannot 
be controlled independently. More recently, Castro {\it et al.} reported the 
first observation of a tunable electronic gap in a BGB through
magneto-transport measurements of micro-mechanically cleaved 
graphene on a SiO$_2$ substrate \cite{Castro06}. By chemically doping the upper
graphene layer with NH$_3$, it was shown that the BGB
behaves as a  semiconductor with a tunable electronic gap that can be changed 
from zero to a value as large as $0.3$ eV by using fields of 
$\simleq \, 1 \, \text{V/nm}$
(below the electric breakdown of SiO$_2$). The importance of controlling both
the chemical potential and the value of the electronic gap in a semiconductor
cannot be overstated since it can open doors for a large number of
applications, from transistors \cite{Nilsson06_BBB} to photo-detectors
and lasers tunable by the electric field effect. 

In order to understand and control the electronic properties of the
BGB it is important to understand the effects of the unavoidable disorder. 
In this letter we show that bound states exist for arbitrary weak impurity
potentials, and that their properties, such as binding energies and 
localization lengths, can be externally controlled with a gate bias.
Moreover, we obtain the wave-functions of the mid-gap states, 
from which we derive a simple criterion for when the overlap between
wavefunctions becomes important. This overlap results in band
gap renormalization and possibly band tails extending into the gap
region, as in the case of ordinary heavily doped semi-conductors
\cite{Mieghem92}, or impurity bands for deep 
impurities. Unlike ordinary semiconductors, the electronic density of states
can be completely controlled via the electric field effect. 
The impurity interaction problem is studied within the coherent potential
approximation (CPA). 

{\it The Model.}
The low-energy effective bilayer Hamiltonian has the form 
\cite{McCann06,Paco06a} (we use units such that $\hbar = 1$): 
\begin{equation}
  \label{eq:Hkin0}
  \Hca_0 (\vk) =
  \begin{pmatrix}
    \sigma^{*} \cdot \vk + {V/2}
    & \tp (1+\sigma_z)/2 \\
    \tp(1+\sigma_z)/2 & 
    \sigma \cdot \vk -{V/2}
  \end{pmatrix},
\end{equation}
where $\vk = (k_x ,k_y)$ is the 2D momentum measured relative to
the K-point in the Brillouin zone (BZ), $V$ is the potential energy 
difference between the two planes, $t_{\perp} \approx 0.35 \, \text{eV}$
is the inter-layer hopping energy, and $\sigma_i$ ($i=x,y,z$) are Pauli
matrices. We choose units such that the Fermi-Dirac velocity, 
$\vf$, is set to unity 
($\vf = 3 t a/2$ where $t \approx 3 \,\text{eV}$ 
is the nearest neighbor hopping
energy, and $a \approx 1.4 \, \text{\AA}$ is the lattice spacing).
The corresponding spinor has weight on the different sublattices
according to
$\Psi = (\psi_{\rA 1}, \, \psi_{\rB 1}, \, \psi_{\rA 2}, \,
\psi_{\rB 2})^{T}$ \cite{Nilsson06a_short}.
Solving for the spectrum of (\ref{eq:Hkin0}) one finds two pairs of
electron-hole symmetric bands:
\begin{equation}
  \label{eq:eigenvalues}
  E_{\pm \pm}^2 = 
  k^2 + V^2 / 4 + \tp^2 /2 
  \pm \sqrt{( V^2 +\tp^2) k^2 + \tp^4 / 4 }.
\end{equation}
For the two bands closest to zero energy near the band edge (valence and
conduction bands) one can write the spectrum as:
$E_{\pm -}(k) \approx \pm \bigl[\Eg/2 + (k-\kg)^2 / (2 \mg) \bigr]$.
Here $\Eg = V \tp / \sqrt{ V^2 + \tp^2}$ is the energy gap,
$\kg = {V/2} \sqrt{1+ \tp^2 /( V^2 + \tp^2 ) }$
is a momentum shift, and $\mg$ is an effective mass.
Due to the non-zero $\kg$ the system is effectively one-dimensional
(1D) near the band edge \cite{Paco06a}.

\begin{figure}[htb]
\includegraphics[scale=0.44]{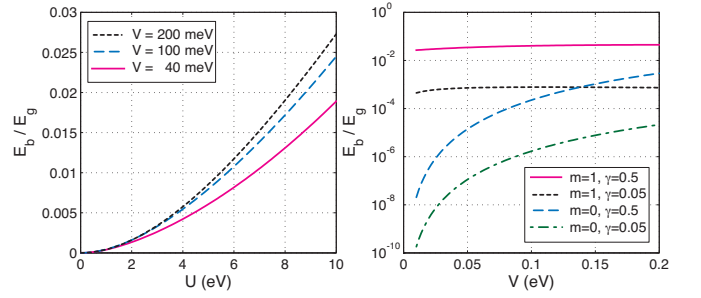}
\caption{(Color online) Left: Bound state binding energies $\Eb$ 
(in units of the gap $\Eg$) for a Dirac delta potential 
%as a function 
of strength $-U$ for different bias $V$. 
Right: Binding energy of a potential well of range $R=10 \, a$ and strength
  $\gamma = \gamma_1 = \gamma_2$ (see text) for different angular momentum $m$.
%, and the external potential $V$. 
}
\label{fig:boundstates1}
\end{figure}

{\it Dirac delta potentials.}
True bound states must lie inside of the gap so that their energies fulfill
$|\e| < \Eg/2 $.
For a single local impurity the Green's function can be written as
$  \GR = \GR^{0} + \GR^{0} T \GR^{0}$ using the standard t-matrix approach 
\cite{JonesMarch2}, where
the bare Green's function is $ \GR^{0} = (\om - \Hca_0)^{-1}$.
The Fourier transform of a Dirac delta potential is $U / N$ 
($N$ being the number of unit cells) leading to
$ T = (U/N)/(1-U \overline{\GR}_{\alpha j }^{0})$.
Here $\alpha =$ (A, B) and $j =$ ($1$, $2$) 
label the lattice site of the impurity.
The quantity
$  \overline{\GR}_{\alpha j }^{0}(\om) 
= \sum_{\vk} \GR_{\alpha j \, \alpha j}^{0}(\om,\vk)/N$
is the local propagator at the impurity site
and the momentum sum is over the whole BZ.
Bound states are then identified by the additional
poles of the Green's function due to
the potential, these are given by 
$\overline{\GR}_{\alpha j }^{0}(\e) = 1/U$.
For energies inside of the gap we find:
\vspace{-0.2cm}
\begin{multline}
  \label{eq:Gbars0}
  \overline{\GR}_{\rA 1}^{0}  = \frac{ V/2 - \om}{2 \Lambda^2} 
  \Bigl\{
  \log \Bigl( \frac{\Lambda^4}{M^4 + ( V^2 / 4 + \om^2)^2} \Bigr) 
\\
 -  \frac{ 2 \om V}{M^2}
  \Bigl[ \tan^{-1} \Bigl( \frac{ V^2 / 4 + \om^2}{M^2} \Bigr) +
  \tan^{-1} \Bigl(\frac{\Lambda^2}{M^2} \Bigr) \Bigr]
  \Bigr\},
\end{multline}
\vspace{-0.6cm}
\begin{multline*}
  \overline{\GR}_{\rB 1}^{0}  = \overline{\GR}_{\rA 1}^{0} 
  \\
  - \frac{({V/2} + \om) \tp^2}{ M^2 \Lambda^2} 
  \Bigl[ \tan^{-1} \Bigl( \frac{ V^2 / 4 + \om^2}{M^2} \Bigr) +
  \tan^{-1} \Bigl(\frac{\Lambda^2}{M^2} \Bigr) \Bigr],
\end{multline*}
where $M^2 =  \sqrt{ V^2 \tp^2 / 4 - \om^2 ( V^2 + \tp^2 )}$,
and $\Lambda$ ($\approx 7$ eV) is a high energy cut-off 
\cite{Nilsson06b_short}.
The corresponding expressions in plane 2 are obtained by the
substitution $ V \rightarrow - V$.
From this we conclude that a Dirac delta potential always 
generates a bound state %(no matter how weak the potential is)
since $\overline{\GR}^{0}$ diverges as the band edge 
is approached (where $M \rightarrow 0$).
The dependence on the cut-off 
(except for the overall scale) is rather weak 
so that the linear in-plane approximation to the spectrum
should be a good approximation as in the case of
graphene \cite{Wehling1short}.
For a given strength of the potential $U$, there are four different
bound state energies depending on which lattice site it is sitting on.
In Fig.~\ref{fig:boundstates1} we show the binding energy as a function 
of $U$ and $V$ 
for the deepest bound state. % (coming from $\overline{\GR}_{\rB 1}^{0}$).
In the limit of $U \rightarrow \infty$ the particle-hole symmetry of
the bound state energies is restored.
%
%
%For illustrative purposes 
We will only consider attractive potentials
in this work, analogous results will hold for repulsive potentials
because of the particle-hole symmetry of the model.
For smaller values of the potential ($|U| \ll \Lambda$) the binding
energy measured from the band edge: $\Eb = \Eg/2 - \e$ 
grows as $U^2$ and the states are weakly bound.
For example, for $ V = 40 \, \text{meV}$ and
$U \lesssim 1 \text{eV}$ one finds
$\Eb \lesssim 4 \times 10^{-4} \Eg$.

{\it Angular momentum states.}
For any potential with cylindrical symmetry it is useful to
classify the eigenstates according to their
angular momentum $m$.
The real-space version of (\ref{eq:Hkin0}) in polar coordinates is
obtained by the substitutions
$k_x \pm i k_y \rightarrow -i (\PD_{x} \pm i \PD_y) 
\rightarrow -i e^{\pm i \varphi}(\PD_r \pm i \PD_{\varphi} /r)$.
Adding a potential that only depends on the radial
coordinate $r$
one can (in analogy with the usual Dirac equation \cite{mele})
construct an angular momentum operator that commutes with the
Hamiltonian.
The angular ($\varphi$) dependence of the angular momentum $m$ eigenstates
are those of the vector:
  \begin{equation}
    \label{eq:vecm}
     u_{\alpha,m}(\varphi) = %\frac{e^{i l \varphi}}{\sqrt{2 \pi}}
     e^{i m \varphi}
  \bigl[
     1            , \,
     e^{-i \varphi} e^{-i \alpha \pi/2}  , \,
     1            , \,
     e^{i \varphi} e^{i \alpha \pi/2}
    \bigr]^T .
  \end{equation}
The parameter $\alpha$ is introduced for later convenience.
If the potential generates bound states inside of the gap these states
decay exponentially: $\sim r^{\gamma} e^{- \kappa r}$.
Assuming that the potential decays fast enough,
the asymptotic behavior of 
(\ref{eq:Hkin0}) implies that the allowed values of $\kappa$ are
\begin{equation}
  \label{eq:kplusminus}
  \kappa_{\pm} = \sqrt{-(\e^2+ V^2 / 4) \pm i M^2 },
\end{equation}
so that weakly bound states have a localization length
\begin{equation}
  \label{eq:lloc}
  l \sim \bigl[2 \kg / (V \tp) \bigr] \sqrt{\Eg / \Eb},
% / \sqrt{(1/2 - |E_b| / \Eg)},
\end{equation}
that diverges as the band edge is approached and decreases 
with %the increase of the 
increasing bias voltage.

{\it Free particle wave-functions.}
The free particle wave-functions in the angular momentum basis can be
conveniently expressed in terms of the following vectors:
  \begin{equation}
    \label{eq:vecW}
     v_{Z,m}(z) =
  \bigl[
    Z_m(z)      , \,
    Z_{m-1}(z) , \,
    Z_{m}(z)    , \,
    Z_{m+1}(z)
    \bigr]^T ,
  \end{equation}
  \begin{equation}
    \label{eq:vecp}
      w(p) =
  \begin{pmatrix}
    \bigl[ (\om + {V/2})^2 - p^2 \bigr] (\om - {V/2}) \\
    \bigl[ (\om + {V/2})^2 - p^2 \bigr] p \\ 
    \tp (\om^2 - V^2 / 4 ) \\ 
    \tp (\om - {V/2}) p
  \end{pmatrix}.
  \end{equation}
The last vector is useful as long as $\om \neq {V/2}$.
The function determining the eigenstates is:
$
%\begin{equation}
%  \label{eq:28}
  D(p,\om) = \bigl[ p^2 - V^2 / 4 - \om^2 \bigr]^2
 + V^2 \tp^2 / 4 - \om^2 ( V^2 + \tp^2 ).
%\end{equation}
$
Then, provided that $D(k,\om) = 0$, ($k>0$) which corresponds to
propagating modes, the eigenfunctions are proportional to
$
\Psi_{Z,m}(\om,k,r) = u_{1,m}  \star v_{Z,m}(k r) \star w(k),
$
where $Z_{m}(z) = J_{m}(z)$ or $Y_{m}(z)$ are Bessel functions.
The star product of two vectors is a vector with components defined by 
$[a \star b ]_{\alpha j} = a_{\alpha j} b_{\alpha j}$.
If on the other hand $D(i \kappa, \om) = 0$, ($\text{Re}[\kappa] >0$) 
the eigenfunctions are:
  \begin{eqnarray}
  \label{eq:PsiKI}
  \Psi_{K,m}(\om,\kappa, r) &=& 
  u_{0,m} 
  \star v_{K,m}(\kappa r)
  \star w(i \kappa ) ,
\\
  \nonumber
  \Psi_{I,m}(\om,\kappa,r ) &=&
  u_{0,m} 
  \star v_{I,m}(\kappa r)
  \star w(-i \kappa ) ,
\end{eqnarray}
with $I_{m}(z)$ and $K_{m}(z)$ being modified Bessel functions.
That these vectors are eigenstates can be verified
directly by applying the real space version of
(\ref{eq:Hkin0}) to them.

{\it Local impurity wave-functions.}
The bound state wave-functions can be read off from  the t-matrix
equation. For a Dirac delta potential the resulting wave-functions
are proportional to the propagator from the site  of the impurity to
the site of interest evaluated at the energy of the bound state. 
%This quantity is also important for studying impurity clusters. 
These propagators are easily expressed in terms of the
free-particle wavefunctions, e.g.:
\begin{eqnarray}
  \label{eq:PsiA1}
  \GR_{\alpha j,\rA 1} &=& \Bigr[
  \frac{ \Psi_{K,0}(\e,\kappa_{+}, r) 
       - \Psi_{K,0}(\e,\kappa_{-}, r) }
       {-i 2 M^2 \Lambda^2}  
  \Bigr]_{\alpha j},
\\
  \nonumber
  \GR_{\alpha j,\rB 1} &=& \Bigl[
  \frac{ \kappa_{+} \Psi_{K,1}(\e,\kappa_{+}, r) 
         - \kappa_{-} \Psi_{K,1}(\e,\kappa_{-}, r) }
       {2 M^2 \Lambda^2 ({V/2}-\e)}
       \Bigr]_{\alpha j}.
\end{eqnarray}
There are also analogous contributions coming from the other valley.
The asymptotic behavior of the modified Bessel functions
$K_{n}(z) \sim \exp(-z)/\sqrt{z}$ as $z \rightarrow \infty$
implies that the bound states %of local potentials
are exponentially localized on a scale that is the same as in the
general consideration above %leading to 
in (\ref{eq:kplusminus}) and (\ref{eq:lloc}).
%This is of course no coincidence.
At short distances one may use that 
$K_{n}(z) \sim 1/z^n$ for $n \geq 1$ 
%and $K_0(z) \sim -\ln z$ 
to conclude that the 
wave-functions are not normalizable in the continuum.
The divergence is however not real since in a proper treatment of the
short-distance physics it is cut-off by the lattice spacing $a$.
The characteristic size of the wave-functions allows for a
simple estimate of the critical density of impurities $n_c$
above which the overlap, and hence the interaction between the %different 
impurities, becomes important. For weakly bound states we estimate
$
n_c \sim [ V \tp / ( \kg t)  ]^2 /( 2 \pi \sqrt{3} ) \, (\Eb / \Eg),
$
indicating that the critical density increases with the applied gate voltage.
For $V \ll t_{\perp}$ we find 
$n_c \approx 2.5 \times 10^{-3} \, (\Eb / \Eg)$,
and for $U \lesssim 1 \, \text{eV}$ one has $n_c \lesssim 10^{-6}$.
%Hence, even small concentrations of impurities lead to wavefunction overlap.
This result shows that even a small amount of impurities can have strong
effects in the electronic properties of the BGB.

{\it Variational approach.} 
A simple variational wave-function consisting of a wave-packet with angular
momentum $m$ and a momentum close to $k_g$ from the $E_{+-}$ band is: 
$
\Psi_{\text{var}}(\xi) \propto \int_{k_g-\xi}^{k_g+\xi} dp 
\Psi_{J,m}(E_{+-}(p),p,r) \sqrt{p / \xi},
$
where we assume that $\xi \ll \kg$.
A variational calculation shows that {\it for any} $m$,
a weak attractive potential of strength $\propto U$ 
leads to a weakly bound state with binding energy
$\Eb \propto U^2$.
This can be understood by noting that
for each value of $m$, due to $\kg$ being non-zero, 
the problem maps into a 1D system with an effective local
potential,
%. It is well-known that 
and
in 1D a weak attractive potential
($\propto U$) always leads to a bound state with binding energy
$\Eb \propto U^2$. Thus the result is a direct consequence of
the peculiar topology of the BGB band edge  \cite{trigonalcomment}.
%The binding energies for large $m$ is usually very small however
%so that the states are probably unobservable.

{\it Potential well.}
The free-particle solutions can be used to study a
simple BGB potential well modeled by two potentials: 
$g_j = -\gamma_j \Theta(R-r)/R$.
$\Theta$ is the Heaviside step function so that
$R$ is the size of the well and $\gamma_j$ its dimensionless 
strength in plane $j$.
Bound states of the potential well are then described by 
two $\Psi_{K,m}$ outside and the appropriate
pair of $\Psi_{J,m}$'s and $\Psi_{I,m}$'s
inside of the well.
By matching the wave-functions at $r=R$ we have studied the 
binding energies and find that the deepest
bound states are in one of the angular momentum channels $m=0,\pm 1$
for a substantial parameter range.
Since these types of states are also present for the Dirac delta potential
we argue that the physics of short-range potentials can be 
approximated (except for the short-distance physics)
by Dirac delta potentials with a strength tuned to give the 
correct binding energy.
A typical result for the binding energies is shown in
Fig.~\ref{fig:boundstates1}.
%A feature of potentials with a finite range is that upon increasing the 
%strength of the potential the binding energies can be made to
%increase until the state merges with the continuum of the
%lower band and becomes a resonance. 
%New bound states also appears at the upper band edge at some point. 
%We expect a similar behavior for a strong Coulomb potential.
The important case of a screened Coulomb potential generally
requires a more sophisticated approach.
Nevertheless, we do not anticipate any qualitative discrepancies
between a potential well and a screened Coulomb potential.
We expect the screening wave-vector to be roughly
proportional to the density of states at the Fermi energy;
and once the range and the strength of the potential have
been estimated a potential well can be used
to estimate the binding energies. 
We also note that the asymptotic behavior in 
(\ref{eq:kplusminus}) is quite general for a decaying potential.
%The interesting case of an  unscreened Coulomb potential is
%beyond the scope of this study.  

{\it Coherent potential approximation.}
As discussed above, for a finite density $n_i$ of impurities 
the bound states interact with each other leading to the possibility
of band gap renormalization and the formation of impurity bands.
A simple theory of these effects is the 
CPA \cite{Soven67,Velicky68}.
In this approximation the disorder is treated as a self-consistent medium
with recovered translational invariance.
The medium is described by a set of four local self-energies which are
allowed to take on different values on all of the inequivalent lattice
sites. % in the problem.
The self-energies are chosen so that on average there is no scattering
in the effective medium. 
%It has been argued that the CPA is the best 
%single-site approximation to the full solution of the problem 
%\cite{Velicky68}.
Explicitly we introduce the diagonal matrix
$
  \Hca_{\Sigma}(\om) =
  \text{Diag}\bigl[\Sigma_{\rA 1} , \, \Sigma_{\rB 1} , \,
  \Sigma_{\rA 2} , \, \Sigma_{\rB 2} \bigr],
$ 
here and in the following we suppress the
frequency-dependence of the self-energies for brevity. 
Then the Green's function matrix is given by
\begin{equation}
  \label{eq:green1}
  \GR^{-1}(\om,\vk) = \om - \Hca_0(\vk) - \Hca_{\Sigma}(\om).
\end{equation}
Finally, following the standard approach to derive the
CPA \cite{Velicky68,JonesMarch2}, 
we obtain the self-consistent equations:
$
\Sigma_{\alpha j} = n_i U / \bigl[ 1-(U-\Sigma_{\alpha j}) 
\overline{\GR}_{\alpha j } \bigr].
$
Using %the explicit form of 
the propagators obtained from
(\ref{eq:green1}) it is straightforward to compute 
$\overline{\text{G}}$ using the same approximations
that lead to (\ref{eq:Gbars0}).
From these equations one can obtain the density of
states (DOS) on the different sublattices:
$\rho_{\alpha j}(\om) = 
-\Ima \overline{\text{G}}_{\alpha j }
(\om + i \delta) / \pi$. In the clean case one finds:
\begin{eqnarray}
\rho_{\rm \rA 1}^0 &=& \biggl|
   \frac{\om - {V/2}}{2 \Lambda^2}\Bigl[ \chi
 - \frac{ \om V (2-\chi) }{\sqrt{( V^2 + \tp^2) \om^2  - V^2
   \tp^2 /4 }}\Bigr] \biggr|,
\nonumber
\\
\rho_{\rm \rB 1}^0 &=& \biggl| \rho_{\rm \rA 1}^0 + 
\frac{\tp^2 ( \om + {V/2})  (2 - \chi) }{2 \Lambda^2 \sqrt{( V^2 +
    \tp^2) \om^2 - V^2 \tp^2 / 4}} \biggr|,
\end{eqnarray}
%\end{subequations}
for $|\om| \geq \Eg/2$.
Here $\chi = (0,1,2)$ for ($|\om|  \leq {V/2} $,  
${V/2} \leq |\om| \leq \sqrt{\tp^2 + V^2 / 4}$,  $\sqrt{\tp^2 +
  V^2 / 4} \leq |\om| $). 
The corresponding quantities in plane 2 are obtained by the
substitution $ V \rightarrow - V$.
Taking the limit $ V \rightarrow 0$ we recover the 
unbiased result of Ref.~\onlinecite{Nilsson06b_short}.
Notice the square-root singularity that starts to appear 
already above $V/2$.
\begin{figure}[htb]
\includegraphics[scale=0.44]{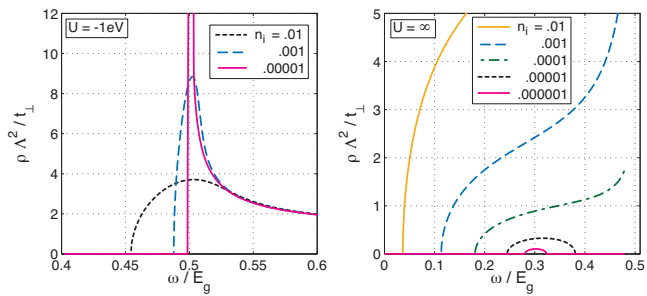}
\caption{(Color online) Left: DOS as a function of the energy 
(in units of $E_g$) close to the conduction band edge 
for different impurity concentrations (see inset),  
%$V = 40$ meV and 
$U=-1$ eV. Right: Details of the DOS 
inside of the gap for different impurity 
concentrations for $U \to \infty$.
In both cases $V = 40 \,\text{meV}$.}
\label{fig:dos1}
\end{figure}

\begin{figure}[htb]
\includegraphics[scale=0.44]{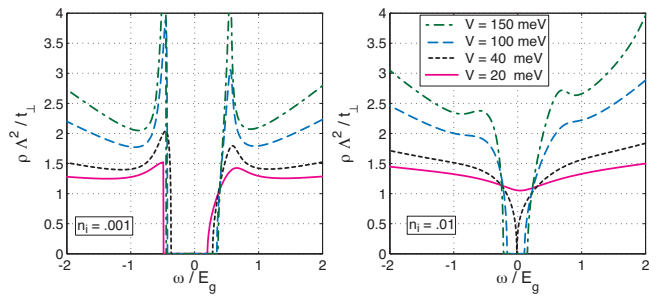}
\caption{(Color online) DOS as a function of energy (in units of $E_g$)
for different values of the applied bias $V$ (see inset) and $U= -10$ eV.
Left: $n_i = 10^{-3}$; Right: $n_i = 10^{-2}$.}
\label{fig:dos2}
\end{figure}

The numerically calculated density of states for $U \to \infty$
is shown in Fig.~\ref{fig:dos1}.
The impurity band evolves from the single-impurity B$2$ bound
state which for the parameters
involved is located at $\e \approx 0.3 \Eg$.
Further evidence for this interpretation is that the total integrated
DOS inside the split-off bands for the two lowest
impurity concentrations is equal to $n_i$.
It is worth mentioning that the width of the impurity band in the
CPA is likely to be overestimated. The reason for this 
is that that the use of effective atoms, all of which have some
impurity character, facilitates the interaction between the impurities 
\cite{Velicky68}.
For smaller values of the impurity strength the single-impurity bound
states are all weakly bound (cf. Fig.~\ref{fig:boundstates1})
and the "impurity bands" merge with the bulk bands
%leading to band gap renormalization, see 
as shown in Fig.~\ref{fig:dos1}
%and the zoom in close to the band edges in 
and \ref{fig:dos2}.
The bands have been shifted rigidly by the amount $n_i U $ for a more
transparent comparison between the different cases. 
The smoothening of the singularity as well as the band gap
renormalization is apparent. Observe also that the band edge moves further into
the gap at the side where the bound states are located.
It is likely that the CPA gives a better approximation for these states
since by (\ref{eq:kplusminus}) they are weakly damped 
almost propagating modes. Notice that the gap and the whole structure of
the DOS in the region of the gap is changing with $V$,
and in particular the possibility that the actual gap closes before $V=0$ 
because of impurity-induced states inside of the gap.

{\it Conclusion.}
We have studied the effects of impurities in a biased graphene
bilayer. We find that local potentials always generate bound states
inside of the gap. 
A finite density of impurities can lead to the formation of impurity
bands as well as band gap renormalization.
We show that, unlike ordinary semiconductors, 
the electronic properties and density of states can
be controlled externally via an applied voltage bias.

\begin{acknowledgments}
We thank A. Geim, F. Guinea and N.~M.~R. Peres 
for many illuminating discussions.
A.H.C.N. is supported through NSF grant DMR-0343790.
\end{acknowledgments}

\bibliography{graphite_0610.bib}

\begin{thebibliography}{23}
\bibitem[{tri()}]{trigonalcomment}
\bibinfo{note}{This feature is unstable to the perturbation of the trigonal
  warping term $\gamma_3$ [6]. The effect of $\gamma_3$ is large for small bias
  and leads to an elliptic dispersion at the band edge. We do not expect
  $\gamma_3$ to alter our results for strong impurity potentials and large
  impurity concentrations.}
\end{thebibliography}

\end{document}